# Direct Visualization of Thermal Conductivity Suppression Due to Enhanced Phonon Scattering Near Individual Grain Boundaries

*Aditya Sood[1,2,*], Ramez Cheaito[2], Tingyu Bai[3], Heungdong Kwon[2], Yekan Wang[3], Chao Li[3], Luke Yates[4], Thomas Bougher[4], Samuel Graham[4,5], Mehdi Asheghi[2], Mark Goorsky[3], Kenneth E. Goodson[2,*]*

[1]Department of Materials Science and Engineering, Stanford University, Stanford, CA 94305, USA. [2]Department of Mechanical Engineering, Stanford University, Stanford, CA 94305, USA. [3]Department of Materials Science and Engineering, University of California, Los Angeles, CA 91355, USA. [4]George W. Woodruff School of Mechanical Engineering, Georgia Institute of Technology, Atlanta, GA 30332, USA. [5]School of Materials Science and Engineering, Georgia Institute of Technology, Atlanta GA 30332, USA.

[*]Authors to whom correspondence should be addressed:

Aditya Sood (aditsood@stanford.edu), Kenneth E. Goodson (goodson@stanford.edu)

ABSTRACT. Understanding the impact of lattice imperfections on nanoscale thermal transport is crucial for diverse applications ranging from thermal management to energy conversion. Grain boundaries (GBs) are ubiquitous defects in polycrystalline materials, which scatter phonons and reduce thermal conductivity ($\kappa$). Historically, their impact on heat conduction has been studied indirectly through spatially-averaged measurements, that provide little information about phonon transport near a single GB. Here, using spatially-resolved time-domain thermoreflectance (TDTR) measurements in combination with electron backscatter diffraction (EBSD), we make localized measurements of $\kappa$ within few $\mu$m of individual GBs in boron-doped polycrystalline diamond. We observe strongly suppressed thermal transport near GBs, a reduction in $\kappa$ from ~1000 Wm$^{-1}$K$^{-1}$ at the center of large grains to ~400 Wm$^{-1}$K$^{-1}$ in the immediate vicinity of GBs. Furthermore, we show that this reduction in $\kappa$ is measured up to ~ 10 $\mu$m away from a GB. A theoretical model is proposed that captures the local reduction in phonon mean-free-paths due to strongly diffuse phonon scattering at the disordered grain boundaries. Our results provide a new framework for understanding phonon-defect interactions in nanomaterials, with implications for the use of high $\kappa$ polycrystalline materials as heat sinks in electronics thermal management.

For many applications ranging from thermal management[1,2] to thermoelectrics[3,4], it is crucial to understand the role that defects and crystalline imperfections play in impeding heat transport at the nanoscale. Grain boundaries, which are two-dimensional defects separating adjacent crystallites in a polycrystalline material, are some of the most commonly found imperfections in solids. Over the past several years, going back to the 1950s, many studies have examined the impact of phonon scattering at grain boundaries (GBs) on thermal conduction in polycrystalline materials[5–9]. The common approach has been to estimate the *average* impact of GBs, by measuring the *effective* thermal conductivity in samples of varying grain size. While these types of measurements have enabled a useful first-order understanding of GB thermal transport, they provided little direct information about the nature of heat flow in the immediate vicinity of an individual boundary. In particular, there remains little understood about the link between diffuse phonon scattering at disorder-rich GBs, and *local* thermal conductivity within a few mean-free-paths (MFPs) of the GB.

This issue is especially relevant to the case of polycrystalline diamond substrates, which have been proposed as next-generation heat sinks for high heat-flux electronic devices[10–13]. It has long been recognized that although diamond holds immense promise due to its record high thermal conductivity (approaching 3000 $Wm^{-1}K^{-1}$ for an isotopically pure single crystal[14]), realistic applications use material that is synthesized by chemical vapor deposition (CVD). CVD is a nucleation and growth process, which results in the formation of a polycrystalline film with a high density of GBs. Due to the strongly diffuse phonon-GB scattering, previous studies[15–18] have shown that the effective (i.e. spatially averaged) thermal conductivity of CVD diamond can be as low as ~10-100 $Wm^{-1}K^{-1}$. Moreover, the columnar GB structure leads to a thermal conductivity tensor which is highly anisotropic and dependent on film thickness[15,19,20]. With continued device scaling, and the development of large-grained high thermal conductivity CVD substrates, we are entering a regime where device dimensions are comparable to or smaller than characteristic grain sizes in the substrate[21]. In this regime, the assumption of an averaged-out, spatially-homogeneous, effective thermal conductivity is no longer valid. Instead, it is essential to investigate how GBs affect the *local* thermal environment that a device experiences.

In this work, we develop spatially-resolved time-domain thermoreflectance (TDTR) measurements, that allow us to make direct observations of thermal conductivity suppression due to phonon scattering near individual GBs. In samples of large-grained boron-doped CVD diamond (average grain size ~ 25 $\mu$m), we correlate the spatial thermal conductivity measurements with local variations in the underlying microstructure, imaged using electron backscatter diffraction (EBSD) in a scanning electron microscope (SEM). Our measurements reveal that the thermal conductivity decreases significantly near a GB, by nearly ~60 % as compared to the peak conductivity inside a grain. Furthermore, we find that this suppression in thermal conductivity is detected by the TDTR probe over a surprisingly long distance, up to ~ 10 $\mu$m away from the GB. To understand these effects, we propose a new theoretical framework, of a reduced local thermal conductivity within a few phonon mean-free-paths (MFPs) of the GB due to increased boundary scattering and non-equilibrium effects. Finite element (FE) calculations carried out within this framework are in good quantitative agreement with our measurements, and point toward the key role played by diffuse phonon scattering at the disorder-rich GBs.

Polycrystalline diamond samples (thickness ~ 530 $\mu$m) were grown by CVD (see Methods for details of growth process). (220) pole figure X-ray diffraction (XRD) measurements (see Figure S1) show that the sample has a strong (110) out-of-plane orientation, but random in-plane orientation, indicated by the ring at $60^0$ corresponding to {110} planes[22]. Figure S2a shows a plan-view SEM image of the sample, which shows some microstructure, although without clearly visible grain boundaries. To better visualize the grain structure, we performed EBSD measurements in the SEM. Figure 1a shows an EBSD integrated intensity image, and Figure 1b shows a map of the in-plane grain orientation where the GBs are clearly identified as boundaries separating adjacent regions of different colors (see Figure S3a-c for maps corresponding to the other orthogonal orientations). This image was produced by generating a diffraction pattern from the backscattered electrons, and using an algorithm to fit each diffraction pattern to a 3D grain orientation pattern. The out-of-plane map shows that a large fraction of grains has (110) planes oriented parallel to the sample surface, while the in-plane maps show that there is no preferential orientation within the plane, consistent with the XRD data. Grain size analysis (see Figure S3d,e) shows that the average size is ~ 23 $\mu$m.

Thermal conductivity measurements were performed using TDTR, which is a well-established optical pump-probe technique[15,23] (see Methods and Figure S4-7 for details). 2D maps of diamond thermal conductivity ($\kappa_{Diam}$) were generated by raster scanning the sample relative to the laser spot at fixed delay time, while recording at each pixel, the in-phase ($V_{in}$) and out-of-phase ($V_{out}$) voltage signals using the lock-in amplifier. This fixed probe delay time is chosen where the measurement sensitivity to the Al/Diamond thermal boundary conductance (TBC) is near-zero, thereby ensuring that variations in the measured TDTR ratio ($= -V_{in}/V_{out}$) signal are entirely due to spatial variations in $\kappa_{Diam}$ (see Methods and Figure S7). We note that the model used to analyze our data assumes that the local thermal conductivity underneath the laser spot is uniform and isotropic. Any spatial inhomogeneities within the sample because of the resistive effect of GBs are therefore convolved into the extracted values of $\kappa_{Diam}$. Our FE modeling approach described later also addresses this aspect.

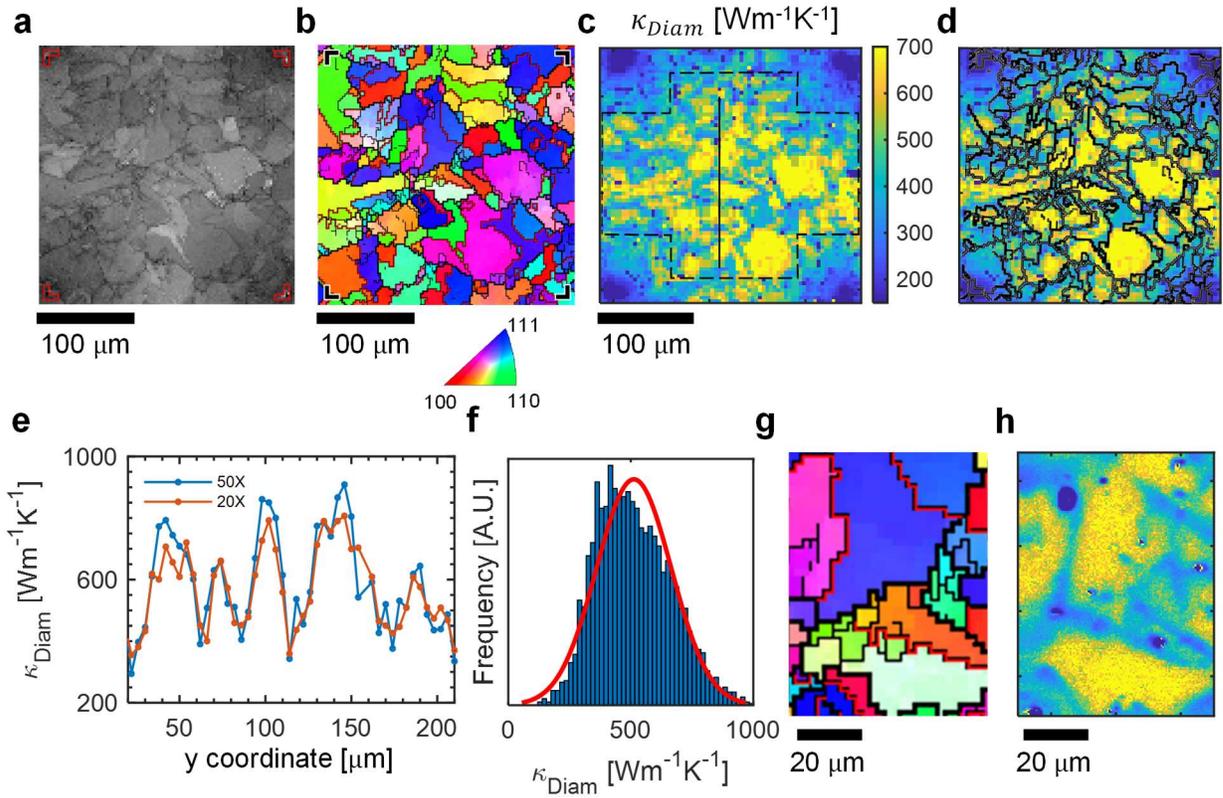

**Figure 1.** Correlative EBSD-TDTR microscopy of polycrystalline diamond. (a) Plan-view EBSD integrated gray-scale image showing the individual grains with GBs. (b) EBSD in-plane grain orientation map, where the color-segment indicates the local crystal orientation relative to the x-axis. (c) TDTR thermal conductivity map (pixel size = 4 $\mu$m, 50x lens with pump/probe diameter = 4.4/3.0 $\mu$m) where high-thermal conductivity is measured inside individual large grains, while GBs have significantly suppressed thermal conductivity. (d) TDTR map overlaid with GBs extracted from the EBSD map, showing a one-to-one correlation between microstructure and local thermal conductivity. (e) Line-scan extracted along the segment marked in (c), showing data taken using both 50x and 20x lenses. (f) Histogram of ~3000 thermal conductivity pixels taken from within the cross-shaped region indicated in (c), along with fit to a normal distribution (mean = 510, standard deviation = 150 Wm$^{-1}$K$^{-1}$). (g) EBSD and (h) high-resolution TDTR maps (step size = 500 nm) of a 60 x 80 $\mu$m region taken from the center of (b) and (c), showing fine microstructural features resolved in the conductivity map.

Figure 1c shows a map of $\kappa_{Diam}$ (step size = 4 $\mu m$, pump/probe spot diameter = 4.4/3.0 $\mu m$) measured over the same region whose SEM and EBSD maps are given in Figure 1a,b. We observe a clear correlation between the local thermal conductivity and grain structure, as illustrated by overlaying the GBs onto the $\kappa_{Diam}$ map in Figure 1d. We measure significantly reduced $\kappa_{Diam}$ (by as much as ~60 % compared to the peak value inside the large grains) for measurements made directly over GBs. This is seen in the line-scan plotted in Figure 1e, where individual grains can be distinguished by the peaks in $\kappa_{Diam}$. In Figure 1f we plot the statistical distribution of pixels taken from the region included within the dashed lines in Figure 1c. In this analysis, we take care to avoid regions close to the FIB cuts, where an artificially lower thermal conductivity is measured, possibly due to defects generated during the ion-milling process. The statistical distribution is non-Gaussian, skewed towards lower $\kappa_{Diam}$ values. This is likely due to a skew in the grain-size distribution (see Figure S3d), similar to the log-normal type grain-size distributions found in materials synthesized by nucleation and growth processes[24]. The statistical analysis of our measurements illustrates important issues regarding the use of TDTR for accurate metrology of heterogeneous materials such as CVD diamond (see discussion in Supporting Note 1). To further push the limits on spatial resolution, we also performed scanning TDTR measurements with step-sizes down to 500 nm (see Figure 1g,h). These measurements (containing ~19000 pixels) show the presence of microstructural features as small as ~6 $\mu$m in the thermal conductivity map.

To better quantify the reduction in $\kappa_{Diam}$ near individual GBs, we look at a region of the sample that contains large grains. Figure 2a-c show the correlative EBSD and TDTR maps of this region, along with line-scans taken across two large grains that are each ~ 40 $\mu m$ across. The peak $\kappa_{Diam}$ is ~1000 Wm$^{-1}$K$^{-1}$, going down to ~400 Wm$^{-1}$K$^{-1}$ at the GBs. In Figure 2d, we plot $\kappa_{Diam}$ versus y-coordinate separation from the GB, for each of the four GBs surrounding the two large grains.

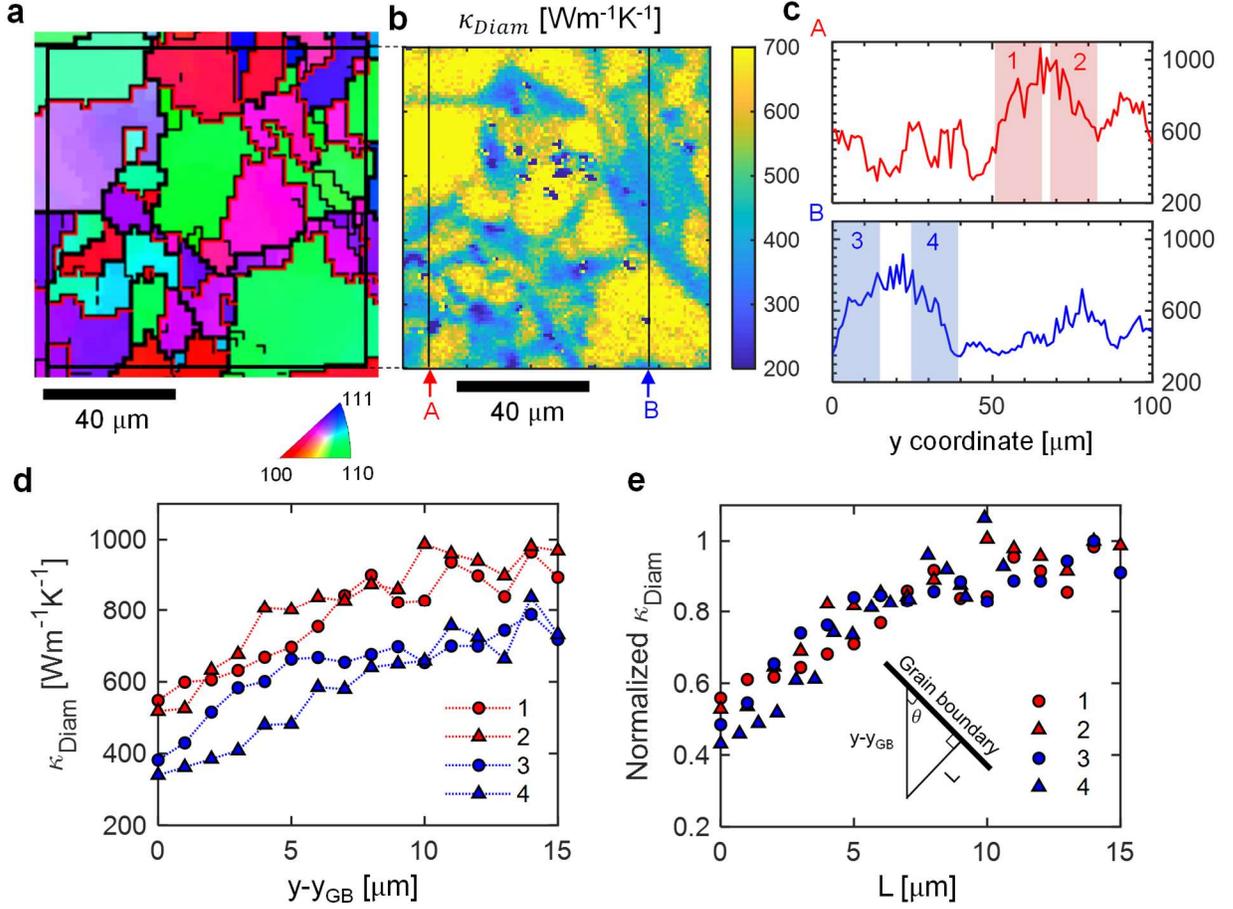

**Figure 2.** Spatially-resolved thermal conductivity measurements near individual grain boundaries. (a) EBSD in-plane grain orientation map, and (b) TDTR thermal conductivity map of a 100 x 100 $\mu$m region (step size = 1 $\mu$m). (c) Line-scans extracted along the segments marked A and B in (b). (d) Thermal conductivity versus y-coordinate measured relative to a GB, from regions marked 1 – 4 in the line scan shown in (c). Data are averaged over a width of $\Delta$x = $\pm$ 3 $\mu$m on either side of the line to reduce noise. (e) Thermal conductivity normalized to its peak value in the center of a grain, versus closest distance to the nearest GB, L. The inset shows how L is calculated; the angle $\theta$ denotes the orientation of the GB in the 2D map. For GBs 1 – 3, $\theta \sim 90^0$ and for GB 4, $\theta \sim 45^0$. Data for all four GBs collapses onto a single curve.

This data is further normalized by the peak $\kappa_{Diam}$ inside each grain, and plotted versus the closest perpendicular distance (L) from the nearest GB, in Figure 2e. We find that the data sets collapse onto a single curve, suggesting a fundamental physical mechanism underlying this reduction in $\kappa_{Diam}$ near a GB. A significant result is that this suppression in thermal conductivity is measured up to relatively large distances from the GB, nearly $\sim 10 \ \mu m$ in each case.

To summarize, our spatially-resolved TDTR measurements reveal two remarkable features: (1) $\kappa_{Diam}$ is reduced by a significant amount near a GB, by almost ~60 % compared to the peak value, and, (2) the resistive effect of a GB is measured by the TDTR probe up to fairly large distances away from it, approaching ~10 $\mu$m. In order to understand the fundamental physical mechanisms responsible for these effects, we develop a modeling framework based on the concept of a spatially varying local thermal conductivity inside a single grain. In general, the local phonon MFP ($\Lambda_{loc}$) at a point located a distance $x$ from a GB (see Figure 3a) can be written using Matthiessen's rule as follows:

$$\Lambda_{loc}(x) = \left[ \Lambda_{bulk}^{-1} + \Lambda_{def}(x)^{-1} + \Lambda_{bdry}(x)^{-1} \right]^{-1} \qquad (1)$$

where $\Lambda_{bulk}$ is the intrinsic MFP due to phonon-phonon scattering in a bulk crystal, $\Lambda_{def}$ corresponds to phonon scattering at point defects such as dopants, and $\Lambda_{bdry}$ corresponds to phonon scattering at GBs. The local thermal conductivity within a grain is related to the local MFP as $\kappa_{loc}(x) = F(x)\kappa_{bulk}$, where $F(x) = \left( \frac{\Lambda_{loc}(x)}{\Lambda_{bulk}} \right)$ is the suppression function.

The second term in Equation 1 captures the effect of spatially-varying phonon-defect scattering rate due to an inhomogeneous distribution of boron dopants. Previous electron energy loss spectroscopy (EELS) studies on boron-doped polycrystalline diamond have shown that the dopant atoms preferentially segregate close to grain boundaries, typically within $\sim$ 1-20 nm of a GB[25,26]. This type of local disorder, also associated with the accumulation of other types of extended defects[27] in CVD-grown diamond, should result in a sharp drop in $\kappa_{loc}$ within a narrow region surrounding a GB. For this case, we model the grain as a homogeneous material with constant conductivity (= 1000 Wm⁻¹K⁻¹), and the GB as a 100 nm thick planar region with reduced conductivity (two different cases, 10 and 100 Wm⁻¹K⁻¹), indicated by the black dotted and dashed lines, respectively, in Figure 3b. These correspond to GB thermal conductance values of $G_{GB}$ = 100 and 1000 MWm⁻²K⁻¹, respectively, which span the range of previously reported estimates in CVD diamond[15,28,29].

On the other hand, the third term in Equation 1 captures the effect of phonon scattering at GBs, and related near-interfacial non-equilibrium effects within a few MFPs of the boundary. To understand the underlying physics behind this non-local effect of a GB, we turn to the Boltzmann transport equation. For phonon transport caused by a temperature gradient along the $z$ direction, we can write an expression for the deviation from equilibrium of the phonon population, $g = f - f_{BE}$, as follows[30]:

$$v \frac{\partial T}{\partial z} \frac{df_{BE}}{dT} + v \frac{\partial g}{\partial x} = -\frac{g}{\tau} \tag{2}$$

where $v$ is the phonon group velocity, $\tau$ is the single-mode relaxation time, and $f_{BE}$ is the Bose-Einstein distribution function. In general, $g$ is a function of both $x$ and $z$. If phonons incident on a GB are absorbed and re-emitted after diffuse scattering according to the local equilibrium distribution function, then $g(0, z) = 0$. The solution can be written as follows[30]:

$$g(x) = -\tau v \frac{\partial T}{\partial z} \frac{df_{BE}}{dT} \left[ 1 - \exp\left( -\frac{x}{\tau v} \right) \right] \tag{3}$$

The deviation function starts from 0 at the GB, and increases to the "bulk" value of $-\tau v \nabla T [df_{BE}/dT]$ on the length scale of the intrinsic phonon MFP $\Lambda_{bulk} = v\tau$. In the gray approximation, the local heat flux along the $z$ direction at a distance $x$ from the GB is proportional to the local departure from equilibrium of the dominant phonon mode, $q_z(x) \propto g(x)$. The local thermal conductivity, given by $\kappa_{loc}(x) \sim q_z(x)/\nabla T$, then has a spatial dependence of the form described by Equation 3.

In essence, we argue that the traditional picture of heat conduction in a polycrystal - where a GB acts as a discrete Kapitza resistance - is incomplete. Instead, a GB exerts its influence non-locally, decreasing the thermal conductivity of the surrounding grains by increasing the rate of phonon scattering. The Boltzmann transport equation suggests that this suppression in local heat flux occurs on a length scale on the order of the dominant phonon MFPs[31]. We note that this is similar to the idea of a surface friction effect, discussed recently in the context of phonon hydrodynamic transport in 2D materials[32].

Moving forward, we model this effect using two simplified expressions. To first order, we assume that for a point located a distance $x$ from a GB, the suppression in MFP is similar to that inside a film of thickness $x$. Using the Matthiessen's approach, the suppression function is as follows:

$$F_M(\delta) = \frac{\delta}{\delta + \left(\dfrac{1-p}{1+p}\right)} \qquad (4)$$

where $p$ is the specularity parameter, denoting the fraction of phonons that are specularly scattered at the GB, and $\delta = x/\Lambda_{bulk}$ (see thick curves in Figure 3b). We also consider a model based on the Fuchs Sondheimer equation[33] (see thin curves in Figure 3b), for which the suppression function is given by:

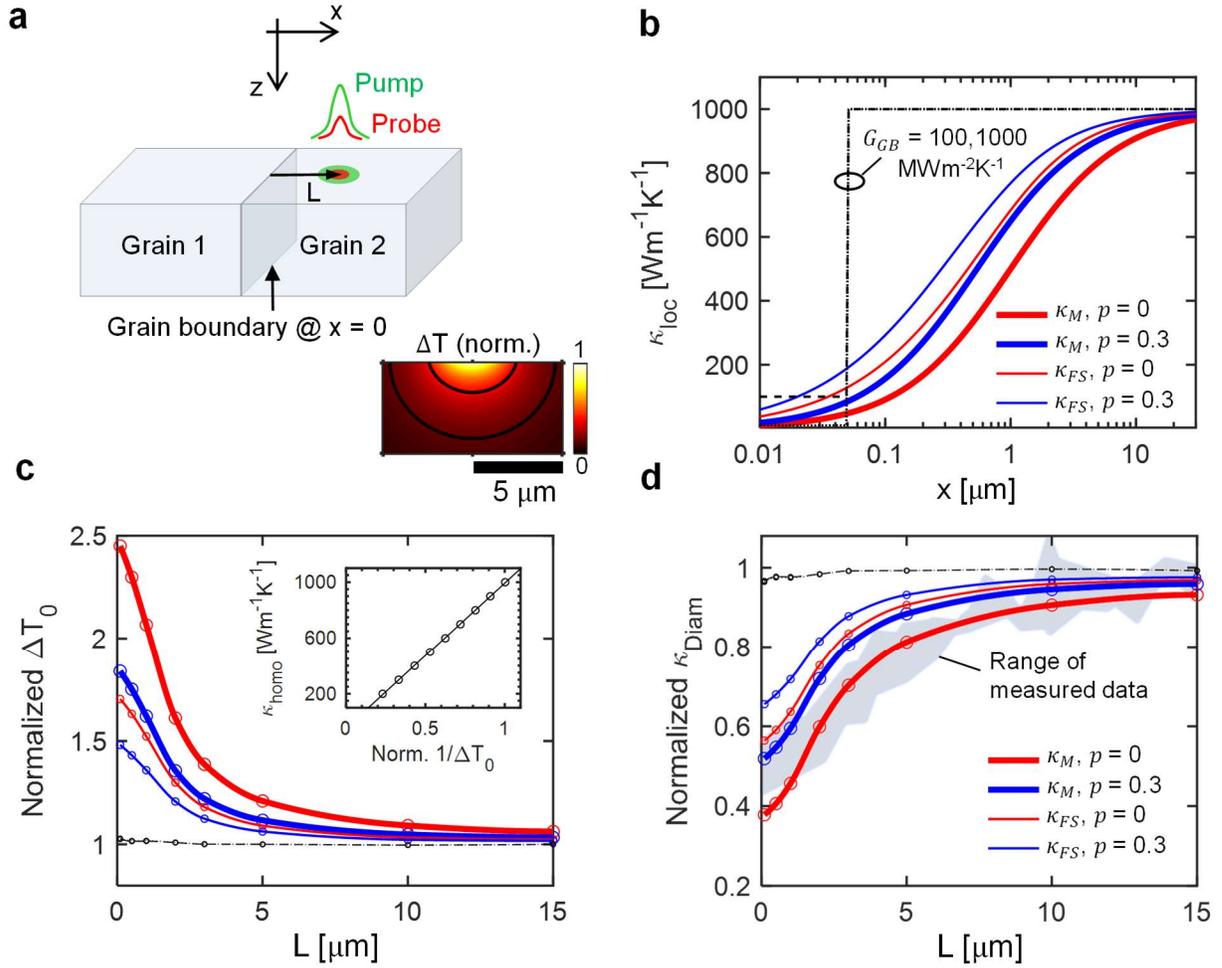

**Figure 3.** Model for thermal transport near a grain boundary. (a) Schematic of finite-element simulation domain showing GB plane located at x = 0. The concentric Gaussian pump/probe spots are displaced by L relative to the GB. Also shown is a plot of the normalized amplitude of temperature oscillations far from a GB, with contours marked at exp(-1) and exp(-2). (b) Local thermal conductivity models: the black dashed (dotted) lines correspond to homogeneous thermal conductivity inside the grain with discrete $G_{GB}$ = 1000 (100) MWm$^{-2}$K$^{-1}$. The thick (thin) red and blue curves correspond to a Matthiessen's type (Fuchs Sondheimer type) model with $p$ = 0 and 0.3, respectively, using a gray phonon MFP of 1 $\mu$m. (c) Amplitude of temperature oscillations averaged over the probe spot versus L, normalized to the bulk value. The inset shows the calibration curve used to convert temperature amplitude to effective thermal conductivity (see Figure S8). (d) Effective thermal conductivity versus L for the different models. For comparison, the gray region indicates the range of measured data from Figure 2e.

$$F_{FS}(\delta) = 1 - \frac{3(1-p)}{2\delta} \int_0^1 (\mu - \mu^3) \frac{1 - \exp\left(-\frac{\delta}{\mu}\right)}{1 - p\exp\left(-\frac{\delta}{\mu}\right)} d\mu \qquad (5)$$

We note that in general, the bulk MFP $\Lambda_{bulk}$ and boundary-limited MFP $\Lambda_{bdry}(x)$ are functions of phonon frequency. For the sake of simplicity, we make a gray medium approximation, i.e. assume a single frequency-independent MFP. Previous density functional theory (DFT) calculations[34] have shown that the MFP spectrum in diamond is significantly narrower as compared to in other semiconductors like silicon, justifying such an approximation[15]; based on these calculations[34], we take $\Lambda_{bulk} \approx 1$ $\mu$m. The parameter $p$ is typically estimated using Ziman's equation: $p \sim \exp(-16\pi^2\eta^2/\lambda^2)$, where $\eta$ is the RMS roughness of the GB and $\lambda$ is the phonon wavelength. Previous studies have shown that there is usually significant disorder associated with CVD diamond GBs [10,27], which would result in highly diffuse phonon scattering ($p$ close to 0).

To translate this $\kappa_{loc}(x)$ into a quantity that can be compared directly with the measured data, we use time-dependent FE simulations. Note that on the short delay times at which the TDTR maps are measured, the $V_{in}$ signal has low sensitivity to $\kappa_{Diam}$, and nearly all our sensitivity comes from $V_{out}$. The out-of-phase component is proportional to the imaginary part of the temperature response of the sample at the modulation frequency $f_{mod}$[35,36]. In our FE simulations, we therefore evaluate the system's temperature response to a sinusoidal heat flux excitation at $f_{mod} = 2.2$ MHz.

The 3D FE simulation domain consists of two grains (40 x 40 x 20 $\mu$m) separated by an interfacial plane at $x = 0$ (see Figure 3a). A 90 nm capping layer simulates the Al transducer, and the interface between the transducer and diamond has a TBC of $G_{Al/Diam} = 100$ MWm$^{-2}$K$^{-1}$. A Gaussian 'pump' heat source (diameter = 4.4 $\mu$m) is applied to the top surface, modulated sinusoidally in time at 2.2 MHz. A concentric 'probe' beam (diameter = 3.0 $\mu$m) measures the Gaussian weighted spatial average of the amplitude of temperature oscillations induced by the pump heat source, $\Delta T_0$. Simulations are performed for different separations (L) of the pump/probe laser spots from the GB. To convert $\Delta T_0(\text{L})$ to $\kappa_{Diam}(\text{L})$, we use the calibration procedure described in Figure S8.

Figure 3c and 3d show plots of the temperature amplitude $\Delta T_0(\text{L})$ and the effective thermal conductivity probed by the laser spot $\kappa_{Diam}(\text{L})$, respectively, for the different thermal conductivity models discussed above. For the first case (see black dashed and dotted lines), where $\kappa_{loc}$ drops sharply near the GBs due to accumulated dopants but is homogeneous elsewhere, the effective thermal conductivity $\kappa_{Diam}$ decreases very slightly (by at most ~5 %) as L approaches 0. This suggests that the experimentally observed suppression in $\kappa_{Diam}$ is unlikely to be caused by the spatial segregation of boron dopants and other defects. On the other hand, for the second case where $\kappa_{loc}$ is gradually reduced due to phonon scattering within a few MFPs of the GB, we find that the simulations predict a significant lowering of $\kappa_{Diam}$ as the laser spots approach the boundary. We calculate a suppression in thermal conductivity of up to ~40 % and ~60 % of the bulk value, for the Fuchs Sondheimer and Matthiessen's models, respectively, in the fully diffuse limit. Furthermore, these models predict that the reduction in thermal conductivity occurs up to ~10-15 $\mu$m away from the GB. For a direct comparison between the simulations and experiments, in Figure 3d we also include the range of measured values, indicated by the gray shaded region. The good agreement between the models and data identifies the important role played by highly diffuse phonon scattering at the disordered GBs, and related near-interfacial non-equilibrium effects, as the physical mechanism likely responsible for the strong suppression in local thermal conductivity.

In conclusion, we have developed a correlative microscopy technique, combining spatially-resolved TDTR and EBSD measurements on bulk samples of boron-doped polycrystalline diamond. These measurements have enabled, to the best of our knowledge, the first direct observations of thermal conductivity suppression in the vicinity of individual grain boundaries. From the point of view of thermal management, our findings have implications for the design of heat-sinking substrates utilizing CVD-grown polycrystalline diamond. They suggest that as device dimensions scale down to length scales comparable to typical grain sizes and phonon MFPs, the local thermal environment can be spatially inhomogeneous depending on the underlying microstructure of the substrate. Furthermore, our work motivates a fresh look at traditional models of thermal transport in polycrystalline materials, where it is commonly assumed that the GBs act as discrete Kapitza resistances. This might be especially interesting to consider in systems where the average grain size is smaller than the dominant phonon MFPs (and hence, the size of the GB influence region)[9,37]. More broadly, the measurement and modeling approach developed here constitutes a new framework for understanding thermal transport in imperfect crystalline materials. Our results are an important advance over research pursued for the past many decades, where the impact of defects on thermal transport was understood in an average sense. We show that high-resolution thermal conductivity measurements near individual defects can provide new insights into fundamental mechanisms of phonon-defect interactions. Similar measurements can be made in other crystalline materials that have intrinsically long phonon MFPs (e.g. silicon and III-Vs) and contain diffusely scattering lattice imperfections. Insights derived from these measurements will enable the rational design of thermal metamaterials with controlled nanoscale architecture, for applications in energy harvesting and thermal management.

**Methods**.

*Sample preparation.* Polycrystalline diamond substrates were grown by microwave enhanced CVD, and heavily doped with boron (~ $10^{21}$ cm$^{-3}$). Source process gases consisted of hydrogen and methane with diborane ($B_2H_6$) as the dopant, and the growth temperature was kept above 1000 $^0$C. Small pieces (~ 13 x 13 mm) were cut from the original wafer, and the top surface was polished down to a root mean square roughness < 5 nm to create a specular reflective surface for TDTR measurements. To help locate the same region of the sample for microstructural characterization (SEM-EBSD) and thermal conductivity measurements (TDTR), fiduciary marks were cut into the corners of a 250 x 250 μm square using a Ga$^+$ FIB. After EBSD measurements in the SEM, the sample was blanket coated with a ~90 nm thick Al transducer layer using electron-beam (e-beam) evaporation, while still allowing the fiduciary marks to remain visible.

*Thermal conductivity measurements.* Details of our TDTR optical setup have been reported previously[15,23]. In these experiments, we used a 50x objective lens with pump and probe laser spot diameters ($1/e^2$) of 4.4 and 3.0 μm respectively (see Figure S5), except for the 20x line scan in Figure 1e where the respective spot diameters were 10.2 and 6.2 μm. The pump beam was amplitude modulated at a frequency of 2.2 MHz. An integrated dark-field microscope enabled sample visualization and location of fiduciary FIB marks under the laser spot (see Figure S4). Full delay time scans were performed at multiple locations within the chosen region to extract the TBC at the Al/diamond interface, giving a value of 112 ± 10 MWm$^{-2}$K$^{-1}$ (see Figure S6). To construct 2D maps of diamond thermal conductivity, we fixed the probe delay time at a value where the signal response is nearly insensitive to the Al/diamond TBC, and raster scanned the sample relative to the laser spot (see Figure S7 for details on how this fixed delay time is calculated). At each pixel, the lock-in amplifier simultaneously records the in-phase and out-of-phase TDTR voltage signals, and DC probe reflectivity. The ratio signal measured at each pixel is converted to thermal conductivity by comparing with a correlation curve obtained from the solution of the multilayer heat diffusion model[35].

ASSOCIATED CONTENT

**Supporting Information**. See the Supplement for additional details of XRD and SEM characterization, grain size analysis, TDTR measurements (sample alignment, spot-size measurements, full time-delay scan results, sensitivity analysis), finite-element simulations, and a note about measurement statistics.

AUTHOR INFORMATION

**Corresponding Authors**

aditsood@stanford.edu, goodson@stanford.edu.

**ORCID**

Aditya Sood: 0000-0002-4319-666X

**Author Contributions**

A.S. conceived the TDTR-EBSD correlative microscopy idea; A.S., R.C., M.A., M.G. and K.E.G designed the experiments; T.B., Y.W. and C.L. performed microstructural characterization and analysis under the supervision of M.G.; A.S. developed the scanning TDTR technique; A.S. performed TDTR measurements with R.C.; A.S. analyzed data with inputs from R.C.; A.S. performed theoretical calculations; H.K. assisted with FE simulations;  L.Y., T.B. and S.G. discussed experimental results; A.S. wrote the manuscript; S.G., M.A., M.G. and K.E.G supervised the project.

**Notes**

The authors declare no competing financial interest.

ACKNOWLEDGMENTS

The authors are grateful to Prof. Avram Bar-Cohen and Prof. Chris Dames for insightful discussions, and Zhe Cheng for comments. This work was supported by the U.S. Defense Advanced Research Projects Agency (DARPA) Microsystems Technology Office (MTO) Diamond Round Robin Program "Thermal Transport in Diamond Thin Films for Electronic Thermal Management" under contract no. FA8650-15-C. The views, opinions, and/or findings contained in this article are those of the authors and should not be interpreted as representing the official views or policies, either expressed or implied of the Defense Advanced Research Projects Agency or Department of Defense. Part of this work was performed at the Stanford Nano Shared Facilities (SNSF), supported by the National Science Foundation under Award No. ECCS-1542152.

# Supporting Information

**X-ray diffraction measurements**

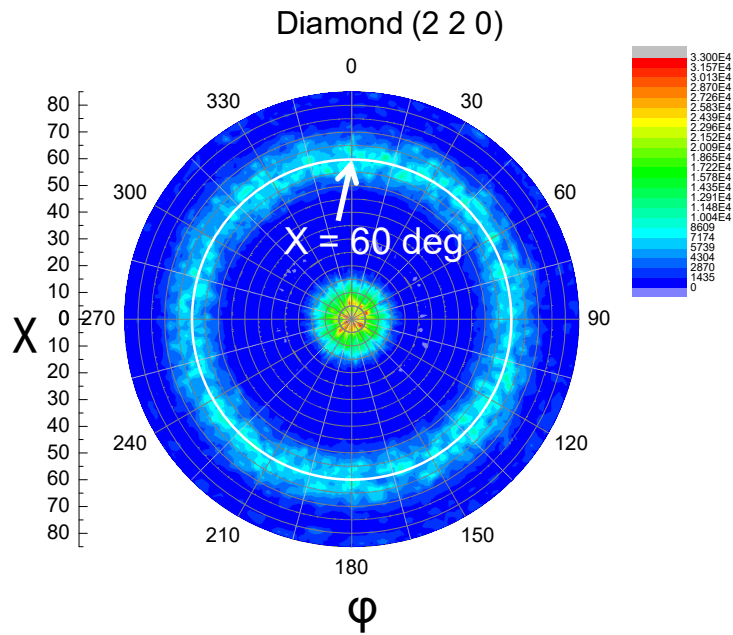

**Figure S1**

(220) X-ray pole figure showing a preferential (110) out-of-plane grain orientation. The ring of lower intensity at 60° corresponds to {110} planes. The fact that this is a continuous ring and not four spots 90° apart demonstrates that the in-plane orientation is random.

**Standard SEM versus EBSD image**

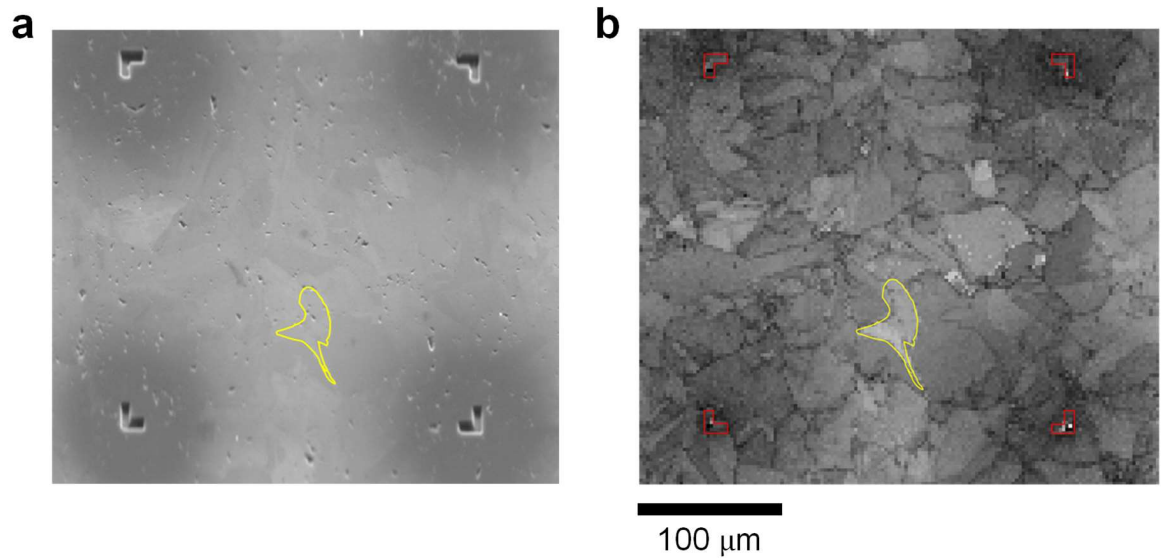

**Figure S2**

(a) Standard plan-view SEM image of a 250 x 250 μm square region showing the corner marks made using FIB. (b) EBSD-rendered STEM-like image where the gray-scale denotes the total integrated intensity of backscattered electrons at each pixel. A single grain is outlined in both images for clarity.

**Grain-size analysis**

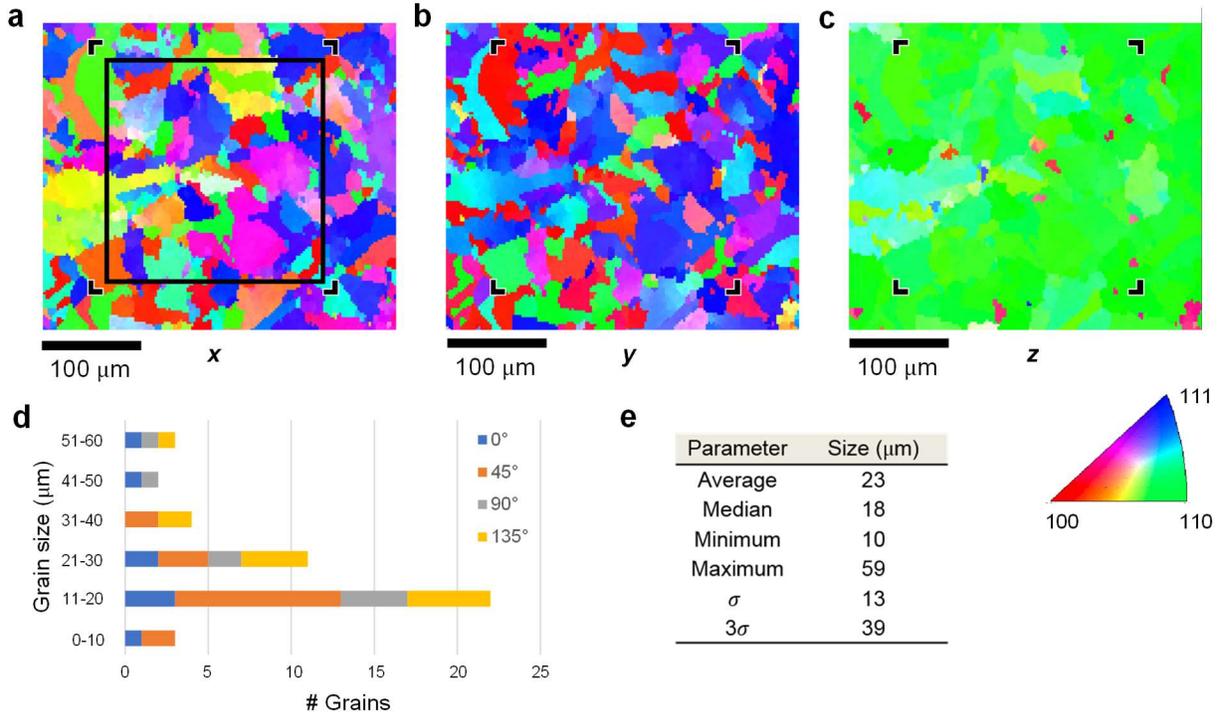

**Figure S3**

(a) In-plane grain orientation map along the x-axis of the figure, where the orientation of individual grains is given by the color-coded segment. (b) In-plane grain orientation along the y-axis. (c) Out-of-plane grain orientation along the z-axis, showing that most grains have a preferential (110) texture, consistent with the X-ray pole figure shown above, while the in-plane orientation is random. (d) Grain-size distribution for the region indicated in (a), and (e) Statistical parameters extracted from this analysis, showing an average grain-size of ~ 23 μm.

**Details of sample alignment procedure**

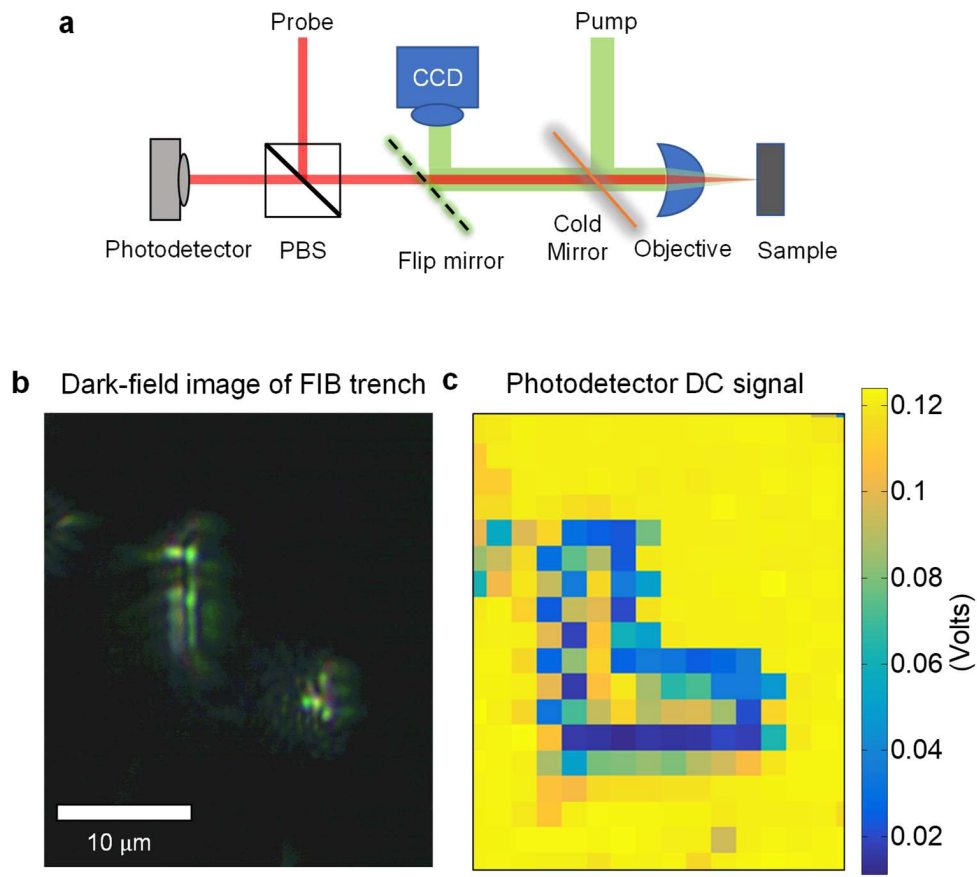

**a**

Probe

CCD

Pump

Photodetector    PBS    Flip mirror    Cold Mirror    Objective    Sample

**b** Dark-field image of FIB trench    **c** Photodetector DC signal

(Volts)

10 µm

**Figure S4**

(a) Schematic of TDTR system showing the location of the charge-coupled device (CCD) camera that allows dark-field imaging of the sample. (b) Dark-field image of one of the L-shaped FIB marks. (c) Map of the DC probe reflectivity of the same region. Sidewalls of the FIB cut trenches scatter light away, leading to a lower signal on the photodetector. This method allows us to pin-point the location of the FIB markers to within ~2 µm, and ensure good registry with the EBSD maps.

**Measurements of the laser spot size.**

We employ the knife-edge method in reflection mode to measure the pump and probe spot sizes. Details of this method are provided in the Supplementary Material of Sood *et al*.[1]

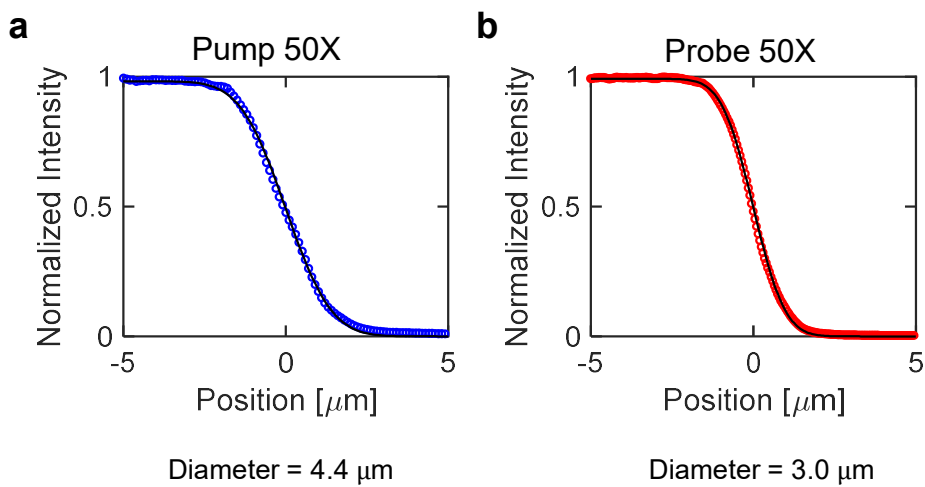

**Figure S5**

Reflected intensity versus laser coordinate, also showing fits to an error function solution. (a) 50x pump, Gaussian diameter = 4.4 μm, (b) 50x probe, Gaussian diameter = 3.0 μm.

**Full delay time TDTR measurements**

Full delay time scans were taken at multiple spots on the sample. The ratio ($= -V_{in}/V_{out}$) data were fit to the solution of a three-dimensional multilayer heat-diffusion model[2,3]. Literature values were used for the volumetric specific heat of Al[4] and diamond[5], $C_{v,Al}$ = 2.43 MJm$^{-3}$K$^{-1}$ and $C_{v,Diam}$ = 1.8 MJm$^{-3}$K$^{-1}$, respectively, and the thickness of the Al transducer was measured using atomic force microscopy (AFM). At each spot, we fit simultaneously for two variables: the thermal conductivity of diamond $\kappa_{Diam}$, and the TBC between the Al transducer and diamond $G_{Al/Diam}$.

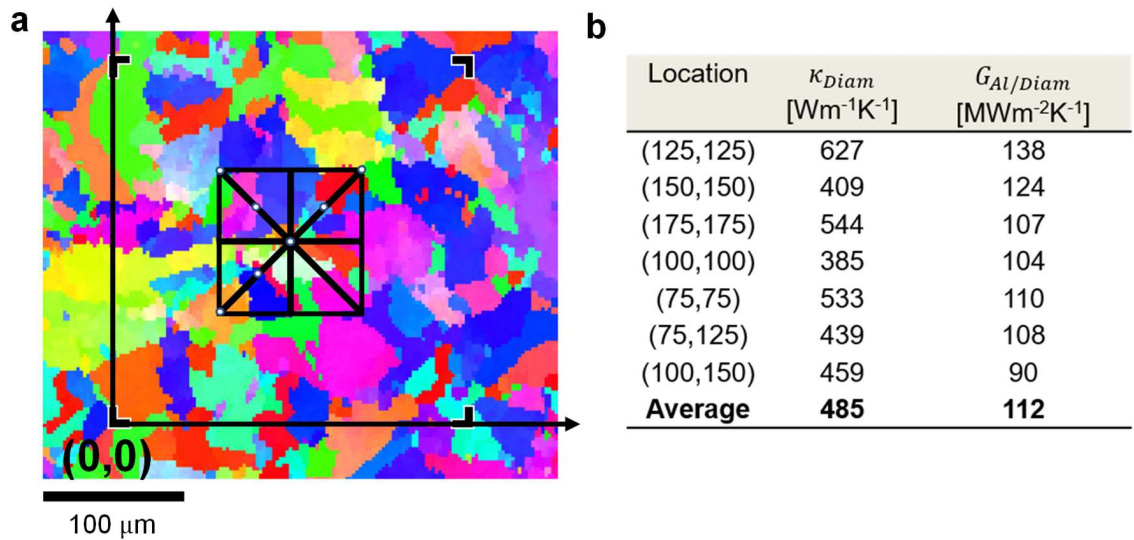

**a**

(0,0)

100 μm

**b**

| Location | $\kappa_{Diam}$ [Wm$^{-1}$K$^{-1}$] | $G_{Al/Diam}$ [MWm$^{-2}$K$^{-1}$] |
|---|---|---|
| (125,125) | 627 | 138 |
| (150,150) | 409 | 124 |
| (175,175) | 544 | 107 |
| (100,100) | 385 | 104 |
| (75,75) | 533 | 110 |
| (75,125) | 439 | 108 |
| (100,150) | 459 | 90 |
| **Average** | **485** | **112** |

**Figure S6**

(a) Location of full delay time TDTR scans marked with white circles, with the origin of the coordinate system defined at the bottom-left FIB marker. (b) Thermal conductivity and Al/diamond TBC values measured using simultaneous two-parameter fitting of full delay time curves at the points indicated in (a). $\kappa_{Diam}$ varies from ~ 385 to 630 Wm$^{-1}$K$^{-1}$, and $G_{Al/Diam}$ varies from ~ 90 to 140 MWm$^{-2}$K$^{-1}$, with no observable correlation between the two quantities.

## Sensitivity analysis

To determine the fixed delay time $\tau_{pb}$ at which the thermal conductivity maps are measured, we calculate the sensitivity coefficients $S_\gamma$ to $\kappa_{Diam}$ and $G_{Al/Diam}$ using the nominal values measured from the full delay time scans. These coefficients quantify the fractional change in the TDTR ratio (= $-V_{in}/V_{out}$) signal due to a fractional perturbation in the parameter $\gamma$ (which is either $\kappa_{Diam}$ or $G_{Al/Diam}$):

$$S_\gamma = \frac{\partial \log\left(-V_{in}/V_{out}\right)}{\partial \log(\gamma)}$$

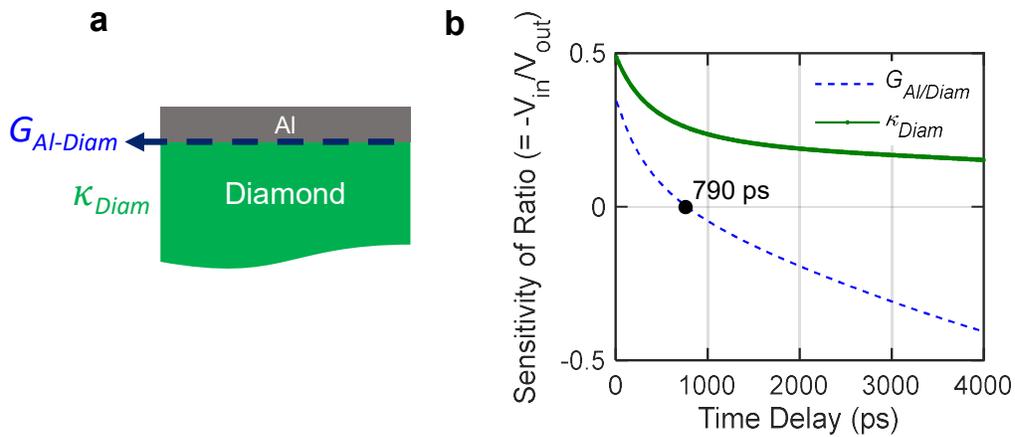

**Figure S7**

(a) Cross-sectional schematic of sample showing the two unknown parameters. (b) Sensitivity coefficients for $\kappa_{Diam}$ and $G_{Al/Diam}$ for the 50x spot size. The measurements have negligible sensitivity to $G_{Al/Diam}$ at $\tau_{pb}$ = 790 ps. All TDTR area maps (using the 50x lens) are taken at this fixed delay time. For the 20x lens (sensitivity coefficients not shown), $\tau_{pb}$ = 300 ps.

**Calibration finite element (FE) simulations on a homogeneous solid**

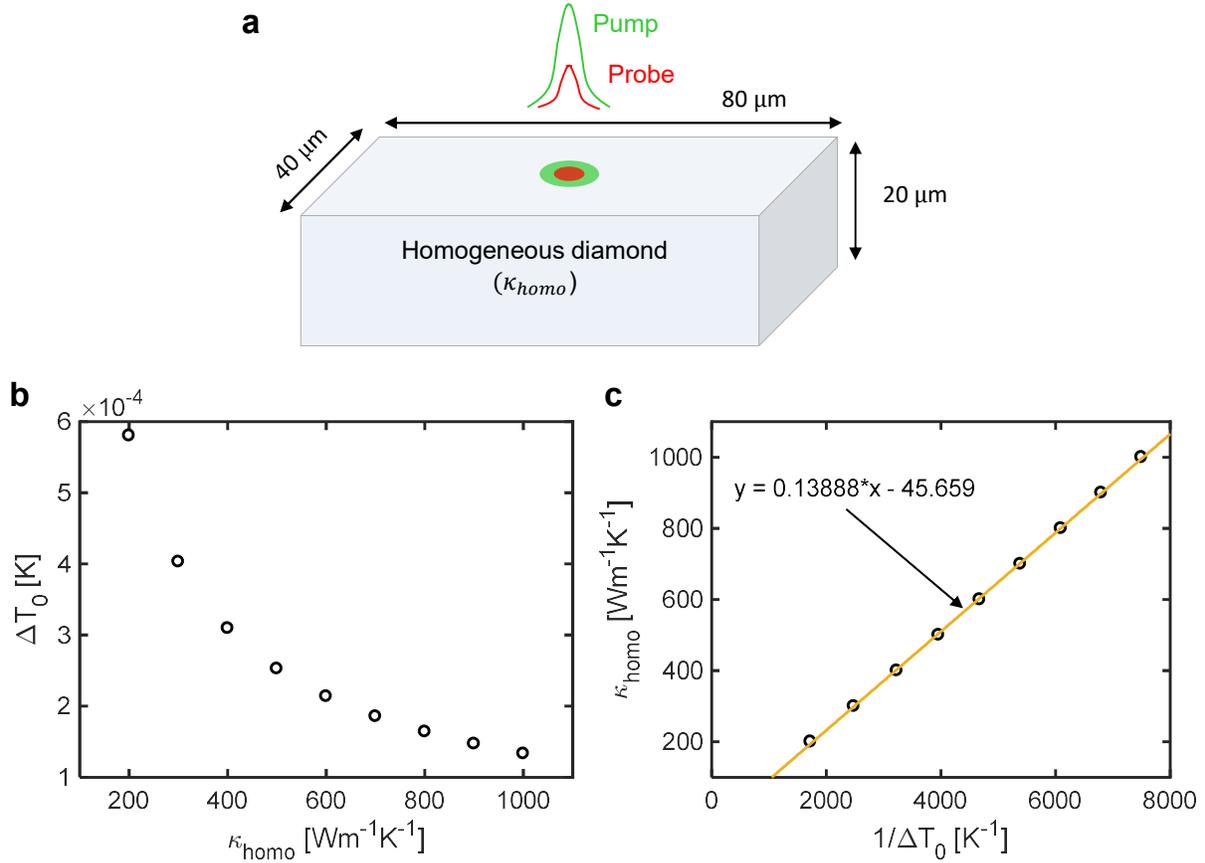

**Figure S8**

Finite-element (FE) simulations of a homogeneous solid (thermal conductivity $\kappa_{homo}$) subjected to an oscillating Gaussian pump heat source. (a) Schematic of FE simulation domain. (b) Amplitude of temperature-rise oscillations averaged over the area of the probe spot, for different values of $\kappa_{homo}$. (c) Plot of $\kappa_{homo}$ versus $1/\Delta T_0$, showing a linear fit. We use this equation to convert the temperature oscillation amplitude in Figure 3c to an effective thermal conductivity shown in Figure 3d, so that the simulations can be compared directly with TDTR measurements.

**Supporting Note 1: Measurement statistics**

The statistical analysis of our spatial mapping measurements illustrates important issues regarding the use of TDTR for accurate metrology of heterogeneous materials. In cases where the length scale of the material heterogeneity (in this case, the grain size) is comparable to or larger than the thermal probe, the locally measured thermal conductivity can vary strongly from location to location. For many applications however, it is important to know the *effective* thermal resistance of the material under a one-dimensional heat flux. This is true for heat-sinking applications where the lateral dimensions of the heat source are larger than the thermal spreading length. In such cases, the effective thermal conductivity that matters in the direction along the applied temperature gradient (say $z$), is simply the mathematical average of conductivity values measured over the $xy$ plane. It is therefore pertinent to address the following question: what is the minimum number of single-spot measurements needed to obtain a reliable estimate of the true, average thermal conductivity of such a heterogeneous material? Intuitively, this should depend on the characteristics of the statistical distribution, i.e. how the standard deviation compares with the mean.

To answer this quantitatively, we invoke the central limit theorem, which states that the minimum number of independent measurements ($n_{MIN}$) needed to ensure that the average lies within ± 10 % of the global mean, with a probability > 95 %, is given by: $2\sigma/\sqrt{n_{MIN}} = 0.1\mu$. Here, $\sigma$ and $\mu$ are the standard deviation and mean, respectively, of the distribution of measured $\kappa_{Diam}$ pixels. Note that the central limit theorem can be applied here even though the starting distribution is non-normal. For our measurements, $\sigma/\mu \sim 0.3$ (see Figure 1e), giving $n_{MIN} \sim 35$. Given that full delay time scans typically take ~2 minutes per spot, this analysis suggests a prohibitively large measurement time using the traditional approach. On the other hand, area maps enable rapid characterization of thousands of points within a relatively short time, with a typical acquisition time of ~ 1 second per pixel.

The measurement technique and statistical approach developed here provides a framework for improved TDTR metrology of spatially heterogeneous samples, where it is important to obtain accurate measurements of average thermal conductivity. Besides polycrystalline materials[6], this approach could be beneficial for the thermal characterization of other defect-rich systems such as radiation-damaged nuclear materials[7], and other technologically relevant composites[8–10].